\documentclass[prb,twocolumn,amsmath,amssymb,showpacs]{revtex4}

\usepackage{graphicx}% Include figure files
\usepackage{dcolumn}% Align table columns on decimal point
\usepackage{bm}% bold math
\usepackage{mathrsfs}
\usepackage{appendix}
\usepackage{bbm}
\usepackage{color}

\def \vS {{\bf S}}
\def \vR {{\bf R}}

\def \vQ {{\bf Q}}
\def \vq {{\bf q}}

\def \vp {{\bf u}}
\def \vv {{\bf v}}

\def \xx {{\bf x}}
\def \yy {{\bf y}}
\def \zz {{\bf z}}

\def \vM {{\bf M}}

\def \vv {{\bf v}}

\def \ee {{\bf e}}

\def \vD {{\bf D}}

\def \vy {{\bf y}}
\def \vx {{\bf x}}
\def \vz {{\bf z}}
\def \qq {{\bf q}}
\def \mB {{\mu_{\rm B}}}
\def \tt {{\bf t}}
\def \br {{\bf r}}
\def \bM {{\bf M}}
\def \blm {{\bf m}}
\def \vD {{\bf D}}
\def \bcr {{\bf m}_{\rm R}}
\def \bci {{\bf m}_{\rm I}}
\def \pp {{\bf u}}

\begin{document}

\title{{Normal Modes of a Spin Cycloid or Helix}\footnote{Copyright notice: This manuscript has been authored by UT-Battelle, LLC under Contract No. DE-AC05-00OR22725 with the U.S. Department of Energy. The United States Government retains and the publisher, by accepting the article for publication, acknowledges that the United States Government retains a non-exclusive, paid-up, irrevocable, world-wide license to publish or reproduce the published form of this manuscript, or allow others to do so, for United States Government purposes. The Department of Energy will provide public access to these results of federally sponsored research in accordance with the DOE Public Access Plan (http://energy.gov/downloads/doe-public-access-plan).}}

\author{Randy S. Fishman$^1$, Toomas R\~o\~om$^2$, and Rog\'erio de Sousa$^3$} 

\affiliation{$^1$Materials Science and Technology Division, Oak Ridge National Laboratory, Oak Ridge, Tennessee, USA}
\affiliation{$^2$National Institute of Chemical Physics and Biophysics, Akadeemia tee 23, 12618 Tallinn, Estonia}
\affiliation{$^3$Department of Physics and Astronomy, University of
Victoria, Victoria, British Columbia, Canada V8W 2Y2}

\date{\today}

\begin{abstract}

Although spin cycloids and helices are quite common, remarkably little is known about the normal modes of a spin cycloid or helix with finite length on a discrete lattice.
Based on simple one-dimensional lattice models, we numerically evaluate the normal modes of a spin cycloid or helix produced by either Dzyaloshinskii-Moriya
(DM) or competing exchange (CE) interactions.  The normal modes depend on the type of interaction and 
on whether the nearest-neighbor exchange is antiferromagnetic (AF) or ferromagnetic (FM).  In the AF/DM and FM/DM cases, there is only a single Goldstone mode;
in the AF/CE and FM/CE cases, there are three.  For FM exchange, the spin oscillations produced by non-Goldstone modes
contain a mixture of tangential and transverse components.  For the DM cases, we compare our numerical results with
analytic results in the continuum limit.  Examples are given of materials that fall into all four cases.

\end{abstract}

\pacs{75.25.+z, 75.30.Ds, 78.30.-j, 75.50.Ee}

\maketitle

\section{Introduction}

Spin cycloids and helices are ubiquitous in the field of magnetism.  They appear in most multiferroics \cite{mostovoy06, cheong07, tok14}
and in many other materials 
like rare earths \cite{elliott61, jensen}, intermetallics \cite{ishikawa76, curro00, nandi09}, and even in some superconductors \cite{lynn97, bao09}.  
Cycloids with spins in the same plane as the ordering wavevector $\vQ $ and helices (also known as spirals or proper screws) with spins perpendicular to $\vQ $ 
partly satisfy neighboring exchange interactions and some competing energy like
Dzyaloshinskii-Moriya (DM) or competing exchange (CE) interactions.   Since DM interactions are usually much weaker
than the nearest-neighbor exchange interactions whereas CE interactions are usually comparable, cycloids or helices produced by DM interactions
typically have much longer periods than those produced by CE.
Cycloids and helices have attracted great attention not only for their accomodating response to competing energies
but also for applications based on their control with electric or magnetic fields \cite{ramesh07}.
Such applications require a deep understanding of the properties of a spin cycloid or helix.

A cycloid with spins in the $xz$ plane propagating along unit vector $\vx $ can be written
\begin{equation}
\vS_r = S\bigl(\sin (Qra), 0 ,\cos (Qra)\bigr),
\end{equation}
where $S$ is the spin and $R=ra$ is the position of site $r$.  With
$\vS_{r+M}=\vS_r$, the magnetic unit cell contains $M$ spins with $0\le r\le M-1$.  
For antiferromagnetic (AF) or ferromagnetic (FM) nearest-neighbor interactions, cycloids with $M=10$ are sketched in Fig.1.
The tangent to the cycloid is given by 
\begin{equation}
{\bf t}_r = \bigl( \cos (2\pi \delta r) , 0 ,-\sin (2\pi \delta r) \bigr),
\label{tngcy}
\end{equation}
where $\vQ=(2\pi /a)(1/2+\delta )\,\vx $ for AF nearest-neighbor coupling and $\vQ=(2\pi /a)\delta \, \vx$ 
for FM nearest-neighbor coupling.
So ${\bf t}_r$ does not alternate sign with the AF modulation in Fig.1(a).

Also known as a spiral or proper screw, a helix with spins in the $yz$ plane
propagating along $\vx $ can be written
\begin{equation}
\vS_r = S\bigl(0,\sin (Qra), \cos (Qra)\bigr)
\end{equation}
with the tangent
\begin{equation}
{\bf t}_r = \bigl( 0,\cos (2\pi \delta r) , -\sin (2\pi \delta r) \bigr).
\label{tnghl}
\end{equation}
Compared to the spin planes of the cycloids in Fig.1, the spin plane of a helix is rotated by $\pi /2$ about $\vz $.
For either a cycloid or a helix, we define $Q_0=Q-2\pi \delta /a$ so that $Q_0=0$ for FM interactions and $Q_0=\pi /a$ for AF interactions.

\begin{figure}
\includegraphics[width=8.5cm]{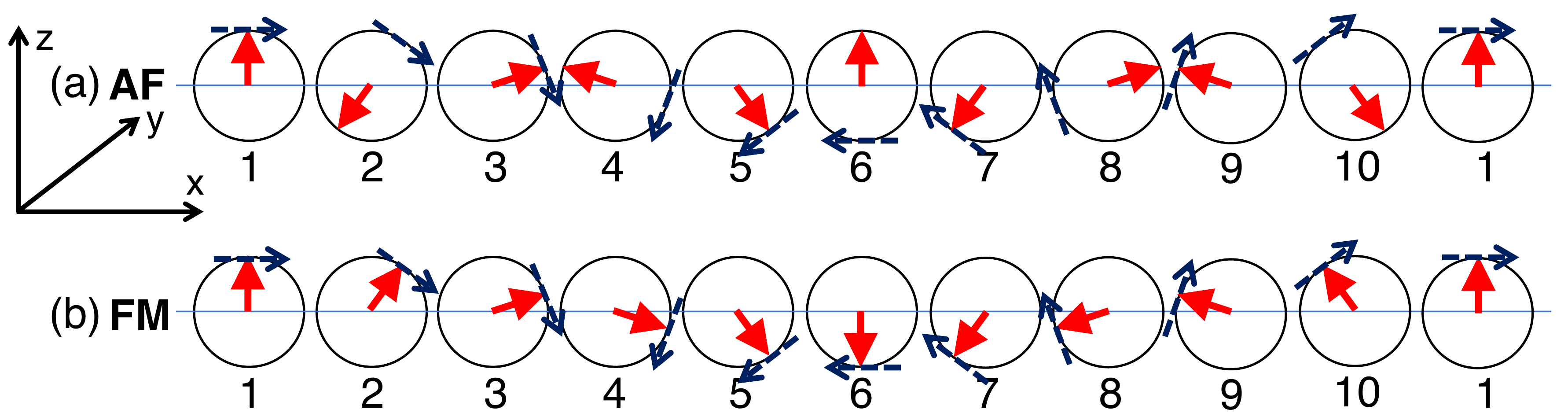}
\caption{(Color online)  Cycloids with (a) AF or (b) FM nearest-neighbor coupling, both with $M=10$.  
Spins $\vS_r$ are solid arrows, tangents ${\bf t}_r$
are dashed arrows.}
\end{figure}

The excitation spectrum of a cycloid or helix provides a dynamical ``fingerprint" of the microscopic interactions and anisotropies
responsible for its formation.
Yet remarkably little is known about the full spectrum of spin-wave (SW) modes for a cycloid or helix,
especially one with a finite period on a discrete lattice.
This paper studies simple one-dimensional lattice Hamiltonians for DM and CE cycloids or helices with either AF or
FM nearest-neighbor exchange.   Our work seeks to answer several questions.   Are
the mode spectra and SW amplitudes different for the four cases (AF/DM, FM/DM, AF/CE, and FM/CE) considered?
Which SW modes can be observed
by inelastic neutron scattering (INS) and which by optical spectroscopy?   When the period is much
larger than the lattice constant, how is the continuum limit (CL)  approached in these four cases?

This paper is divided into seven sections.  Section II describes the simple one-dimensional models for the four cases described
above.  The mode spectra of these modes are solved in Section III.  Section IV describes the method used to obtain
the CL results and the CL
solutions for the AF/DM and FM/DM cases.  Section V evaluates the SW amplitudes for cycloids or helices of finite length in all four cases.
We demonstrate how these SW amplitudes approach the CL.  In Section VI, we discuss the observability of the
SW modes by INS and THz spectroscopy.  Finally, Section VII contains a conclusion.  Details of the CL calculation for the
FM/DM case are provided in the Appendix.

\section{Models}

A one-dimensional lattice Hamiltonian for a DM cycloid or helix is 
\begin{equation}
\label{HDM}
{\cal H}_{\rm DM} = -J_1\sum_i \vS_i \cdot \vS_{i+1}  + D \sum_i \vp \cdot (\vS_{i+1} \times \vS_i),
\end{equation}
where neighboring sites $i$ and $i+1$ are separated by lattice constant $a$ along the $x$ axis.  
For local-moment systems, perturbation theory indicates that the DM vector
$D\vp $ must lie perpendicular to the bond between two spins
so that $\vp $ must lie along $\vy $ or $\vz $ in Eq.~(\ref{HDM}), as first shown by 
Moriya \cite{Moriya60} and Keffer \cite{Keffer62}.
As a consequence, only cycloids would be allowed. 
However, the DM vector may have a  component along the bond direction due to orbital magnetism
\cite{Katsnelson10}  or if magnetic interactions beyond nearest neighbors are taken into account
\cite{Chizhikov13}.

As a starting point, we define 
$\vp = \vy $ for a cycloid and $\vp = -\vx $ for a helix with DM vector $D\vp $.  This definition assures that 
\begin{equation}
\vS_r \times \vp = - S(-1)^{Q_0ra}\,\tt_r,
\end{equation}
\begin{equation}
\vS_r \times \tt_r= S(-1)^{Q_0ra}\, \vp 
\end{equation}
for both cycloids and helices.
For either sign of $J_1$, a cycloid or helix of period $Ma$ is produced by the DM interaction $D$ 
when
\begin{equation}
D = J_1 \tan ( 2\pi  \delta ).
\label{DJ1}
\end{equation}
For AF $J_1 < 0$, $\delta = p/2M$ where $p$ is the number of $2\pi $ rotations (not counting the AF oscillations) in distance $2Ma$.
The cycloid or helix is periodic in distance $Ma$ with $\vS_r = \vS_{r+M}$ if integer $p$ is odd (even) and $M$ is odd (even).  
Otherwise, the AF modulation $(-1)^r$ will give $\vS_r = -\vS_{r+M}$.
For FM $J_1>0$, $\delta = p/M$ where $p$ is the number of $2\pi $ rotations in distance $Ma$.  

A one-dimensional lattice Hamiltonian for a CE cycloid or helix is
\begin{equation}
\label{HCE}
{\cal H}_{\rm CE} = -J_1 \sum_i \vS_i \cdot \vS_{i+1} - J_2 \sum_i \vS_i \cdot \vS_{i+2},
\end{equation}
where $J_2$ is the next-nearest-neighbor exchange coupling
between sites $i$ and $i+2$.  When $\vert J_2\vert $ is sufficiently large, AF exchange $J_2 < 0$ 
frustrates simple AF or FM order to produce a cycloid or helix regardless of the sign of $J_1$.   
The next-nearest-neighbor exchange $J_2$ produces a cycloid or helix with period $Ma$ when
\begin{equation}
J_2 = -\frac{\vert J_1\vert }{4}\sec (2\pi \delta ).
\end{equation}
For $J_1 < 0$, $\delta = p/2M$ and for $J_1 > 0$, $\delta = p/M$ as above.  
Only collinear AF or FM order is possible when
$\vert J_2\vert < \vert J_1\vert/4$.

In ${\cal H}_{\rm DM}$, the classical-spin plane is constrained by the DM interaction 
to lie perpendicular to $\vp $.  But the classical-spin plane is not fixed by the 
CE interactions in ${\cal H}_{\rm CE}$.  The classical-spin plane can then be constrained to lie perpendicular to 
$\vp $ by adding a small (infinitesimal) easy-plane anisotropy energy $-K\sum_i (\vS_i \cdot \vp )^2$ with $K < 0$.  

\section{Mode spectra}

We solve for the SW modes of these two models by performing a $1/S$ expansion 
about the classical limit and then diagonalizing a $2M\times 2M$ equation-of-motion matrix \cite{Haraldsen09, fishmanbook}.  Taking
$S=5/2$, the predicted INS intensities $S(q, \omega )$ are plotted in Fig.2 for all four cases with $\delta =1/10$ ($M=10$ with $p=2$ and
$\delta = p/2M = 1/10$ for 
AF interactions or $p=1$ and $\delta =p/M =1/10$ for FM interactions).   Reciprocal lattice units are defined with $q=2\pi H/a$.  
Clear signatures are exhibited by the spectra of cycloids or helices produced by DM or CE interactions.
For CE cycloids and helices, the SW modes always fall within the first structural Brillouin zone between $H=0$ and $1$, as can be seen 
by using symmetry to mirror the AF/CE
frequencies in Fig.2(c) about $H=0.5$.  For DM cycloids and helices, the SW branches
extend beyond the first Brillouin zone.  For example, three SW branches arise from $H=\pm 1/10 $ and $H=0$ for the
FM/DM case in Fig.2(b).  

\begin{figure}
\includegraphics[width=8.5cm]{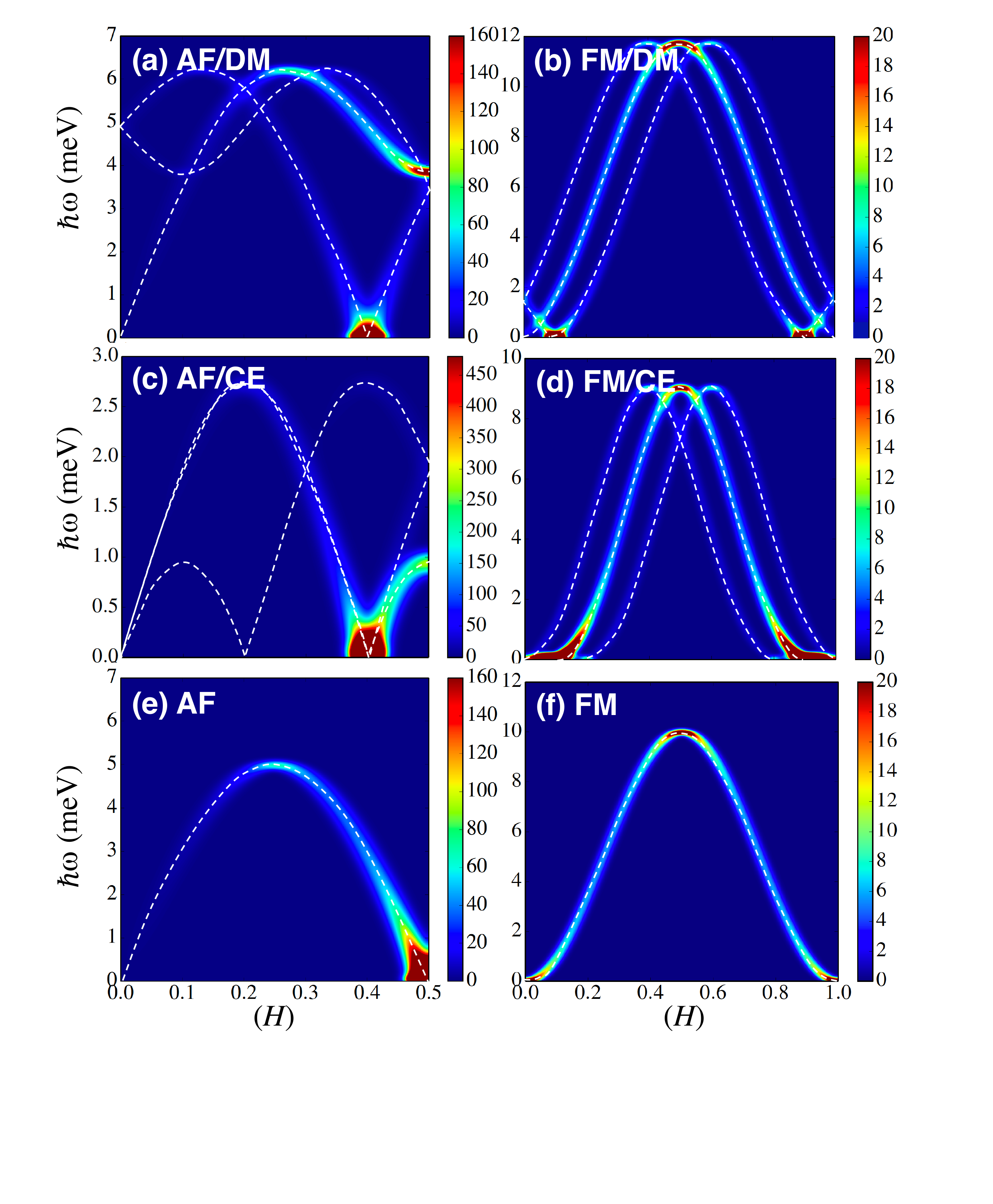}
\caption{(Color online) The INS intensity $S(q, \omega )$ with $q=2\pi H/a$ and $\delta =1/10$
for (a, b) DM and (c, d) CE cycloids or helices with either (a, c) AF or (b, d) FM exchange $\vert J_1\vert = 1$
meV.   For reference, the INS intensities 
for simple AF or FMs with $D=0$ and $J_2=0$ are given in (e) and (f).  Frequencies of modes with significant intensity are drawn as dashed curves.}
\end{figure}

Normal modes evaluated at wavevector $H=m\delta $ (integer $m$) can appear in  
optical measurements since zone folding maps those wavevectors onto $q=0$.
To understand the different mode spectra in our four cases, we plot the SW dispersions versus wavevector $q$ in Fig.3.
Any normal mode crossed by two SW branches is doubly degenerate.  
In Fig.3(a), only one SW branch crosses $\Psi_0$ because the frequencies of $\Phi_{\pm 1}$ are slightly lower than 
that of $\Psi_0$ when $\delta > 0$. 

\begin{figure}
\includegraphics[width=8.5cm]{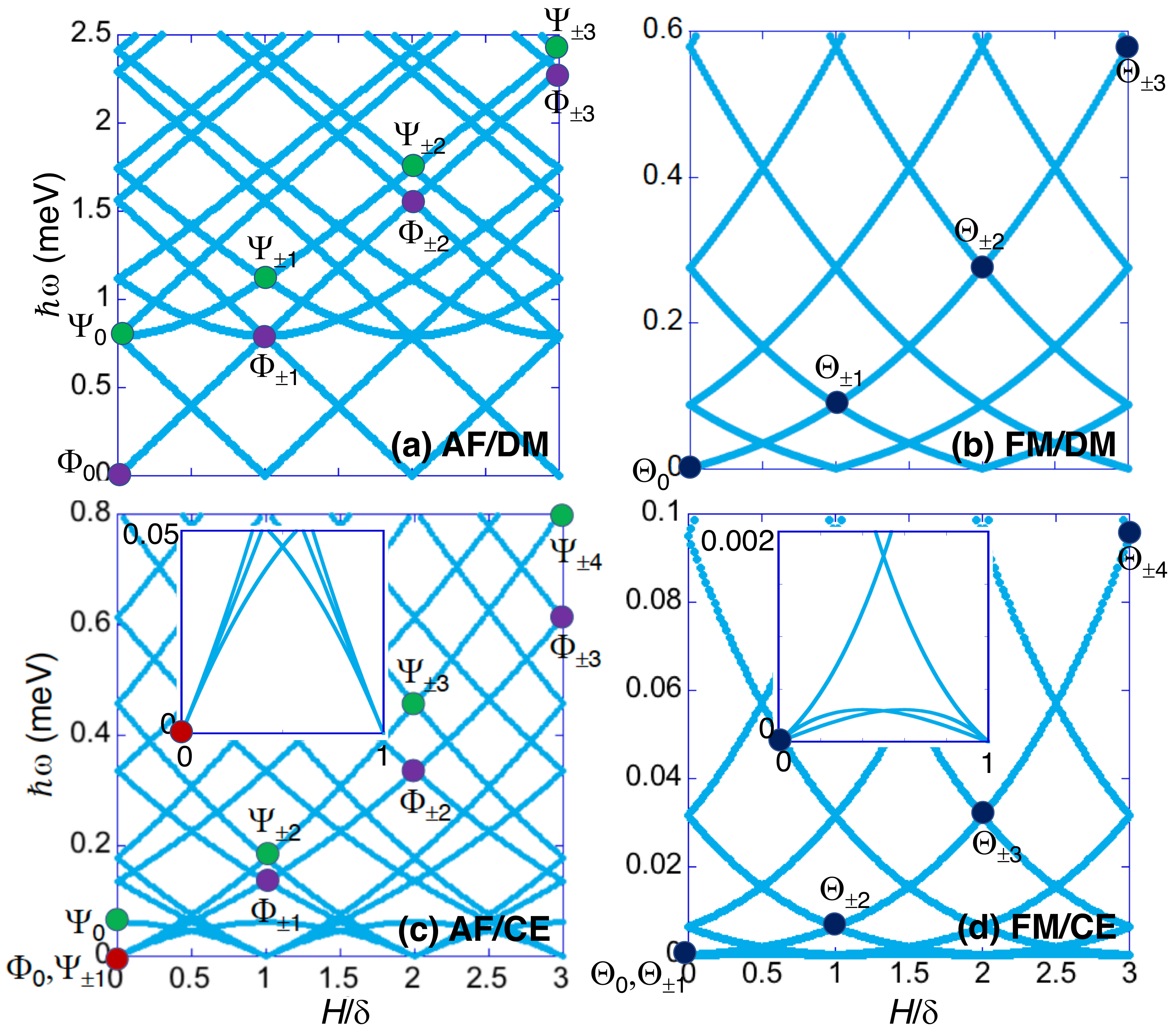}
\caption{(Color online) The SW frequencies versus $H/\delta $ for (a, b) DM and (c, d) CE cycloids or helices with either (a, c) AF or (b, d) FM exchange $\vert J_1\vert = 1$
meV.  In all four cases, $\delta = 1/40$ ($M=40$ with $p=2$ for AF interactions and $p=1$ for FM interactions).
Closed circles are the normal modes discussed in the text.}
\end{figure}

For AF interactions, we obtain two classes of modes labeled $\Phi_{\pm n}$ and $\Psi_{\pm n}$ (doubly degenerate for $n>0$).   
In the AF/DM case, the single Goldstone mode $\Phi_0$ corresponds to a uniform spin rotation about $\vp $.  
In the AF/CE case, the three Goldstone modes are  $\Phi_0$ and 
$\Psi_{\pm 1}$.  Their three-fold splitting away from $H=0$ is plotted in the inset
to Fig.3(c) and can also be seen in Fig.2(c).   Goldstone modes $\Psi_{\pm 1}$ 
are associated with rotations out of the classical-spin plane, assuming that the easy-plane anisotropy $K$ vanishes.  Of course, this
rotation costs energy in the AF/DM case.   

For FM interactions, we obtain only one class of modes labeled $\Theta_{\pm n}$ (doubly degenerate for $n>0$).  
In the FM/DM case, the single Goldstone mode $\Theta_0$ again
corresponds to a uniform spin rotation about $\vp $.  
In the FM/CE case, the three Goldstone modes are  
$\Theta_0$ and $\Theta_{\pm 1}$ with the three-fold splitting plotted in the inset to Fig.3(d).
As in the AF/CE, the extra Goldstone modes are associated with rotations of the spin state out of the classical-spin plane.   
In all four cases, the Goldstone modes are ``massless," meaning that the dispersion is linear near $H=0$.  A linear dispersion along the helical axis was also found by 
Maleyev \cite{Maleyev06} in his study of itinerant cubic magnets.

\section{Continuum limit}

The CL takes $\delta \ll 1$ or $M \gg 1$ so that a period of the cycloid or helix contains
many sites.  Consequently, the spin deviation from one site to the next (neglecting possible AF oscillations) is small.

Previously, de Sousa and Moore \cite{desousa08} found that the CL mode frequencies in the AF/DM case are:
\begin{eqnarray}
\hbar \omega (\Phi_{\pm n}) &=& 2S\vert D n\vert ,\\
\hbar \omega (\Psi_{\pm n}) &=& 2S\vert D\vert \sqrt{1+n^2},
\end{eqnarray} 
where $H = n\delta =n/M$.
As seen in Fig.4(a), the numerical mode spectrum for $\hbar \omega /S\vert D\vert $ 
is close to the predicted spectrum $\hbar \omega /S\vert D\vert = 2\vert n\vert $ or $2\sqrt{1+n^2}$
and the deviations between the numerical and CL results disappear as $M \rightarrow \infty $.
The CL results imply that $\omega (\Phi_{\pm 1})=\omega (\Psi_0) = 2S\vert D\vert $.
For any finite $M$, we find that $\omega (\Phi_{\pm 1}) < \omega (\Psi_0)$ but the difference 
$\omega (\Psi_0)-\omega (\Phi_{\pm 1})$ vanishes as $M \rightarrow \infty $.  

Why do the SW modes for $M=20$ in Fig.4(a) bend towards lower frequencies?  For any finite $M$, the SW frequencies
have zero slope when $H=1/4$ or $H/\delta = M/4$ (see Fig.2(a)).  So for $M=20$, the slope of the SW frequencies
approaches zero as $H/\delta \rightarrow 5$.

We now briefly sketch the CL calculation for the FM/DM case.  Details of this derivation are provided in the Appendix.
The CL is taken for the $d$-dimensional hypercubic Hamiltonian
\begin{equation}
{\cal H}=-J_1 \sum_{\langle i,j\rangle } \vS_i \cdot \vS_j  -\sum_{\langle i,j\rangle }\vD_{ij}\cdot \bigl( \vS_j \times \vS_i \bigr),
\label{htot}
\end{equation}
where nearest-neighbor sites $i$ and $j$ are connected by the vectors ${\bm{v}}=\vR_j -\vR_i$.
In the presence of lattice translation symmetry, the DM vector $\vD_{ij} = \vD_{\bm v}$ 
must be on odd function of $\bm{v}$ so that
$\vD_{\bm v}= -\vD_{-\bm v}$. 
This suggests a simple model with
\begin{equation}
\vD_{\bm{v}}=D_1 \vv + D_2 \left(\ee \times \vv \right).
\label{dvmodel}
\end{equation}
where $\vv ={\bm{v}}/\vert  {\bm{v}} \vert $ is a unit vector.
While $D_1$ can be nonzero only if the lattice breaks inversion symmetry \cite{Chizhikov13}, 
$D_2$ can be nonzero even when the lattice has an inversion center.  In that case, the unit vector $\ee $ would be
parallel to an external or internal electric field, like the ferroelectric moment $\bf{P}$.  
Comparing Eqs.~(\ref{htot}) and (\ref{dvmodel}) to Eq.~(\ref{HDM}), our earlier one-dimensional cycloidal model
follows when $\vv = \xx $, $\ee = \zz $, and $D_2 = -D$.   Our one-dimensional helical model
follows when $\vv =  \xx$ and $D_1 = D$.

With the CL magnetization defined as 
\begin{equation}
\vM (\br )=-g\mu_B \sum_i \vS_i \, \delta(\br-\vR_i),
\end{equation}
the DM Hamiltonian is given by
\begin{eqnarray}
{\cal H}_{DM}&=&-\frac{1}{2}\sum_{i,j,\bm{v}}\vD_{\bm{v}}\cdot \left[ \vS_{j}\times \vS_{i}\right] \delta_{\vR_j,\vR_i+\bm{v}} \nonumber\\
&=& -\frac{V_c}{2(g\mu_B)^2} \int d^d r \int d^d r' \, \sum_{\bm{v}}\vD_{\bm{v}}\nonumber \\
&\cdot &\left[\vM (\br)\times \vM (\br')\right]\,\delta (\br'-\br-\bm{v}),
\end{eqnarray}
where $V_c= a^d$ is the unit cell volume and
the Kronecker delta $\delta_{\vR_j,\vR_i+\bm{v}}$ is replaced by $V_c\, \delta(\vR_j-\vR_i-\bm{v})$ 
in the limit $V_c \rightarrow 0$. 
The DM Hamiltonian can be written as the volume integral ${\cal H}_{DM}=\int d^dr \, h_{DM}$ with density 
\begin{eqnarray}
h_{DM}&=&-\frac{V_c}{2(g\mu_B)^2}\sum_{\bm{v}} \vD_{\bm{v}}\cdot \left[\vM (\br+\bm{v})\times \vM (\br)\right]\nonumber\\
&\approx  &-\frac{V_c}{2(g\mu_B)^2} \, \vM (\br )\cdot \biggl\{  \sum_{\bm{v}}\vD_{\bm{v}}(\bm{v}\cdot \bm{\nabla})\nonumber \\
&\times &\vM (\br) \biggr\}.
\label{hdmaux}
\end{eqnarray}
Using Eq.~(\ref{dvmodel}) to write
\begin{equation}
\sum_{\bm{v}}\vD_{\bm{v}}(\bm{v}\cdot \bm{\nabla}) 
= 2aD_1 \bm{\nabla}+2a D_2 (\ee \times\bm{\nabla}),
\label{sumvDvcubic}
\end{equation}
we then obtain 
\begin{eqnarray}
h_{DM}&=&-D'_{1}\, \vM \cdot \left(\bm{\nabla}\times\vM \right)  -D_{2}' \,
\vM\cdot \Bigl\{ (\ee \times \bm{\nabla})\times\vM \Bigr\}  \nonumber \\ 
&=& -D'_{1}\, \vM\cdot \left(\bm{\nabla}\times\vM \right) 
+D'_{2}\, \ee \cdot\Bigl\{ \vM \left(\bm{\nabla}\cdot\vM \right) \nonumber \\
&+& \vM \times \left(\bm{\nabla}\times \vM \right) \Bigr\} 
\label{finalhdm}
\end{eqnarray}
with $D'_i=a^{d+1}D_i/(g\mu_B)^2$. 

\begin{figure}
\includegraphics[width=7.5cm]{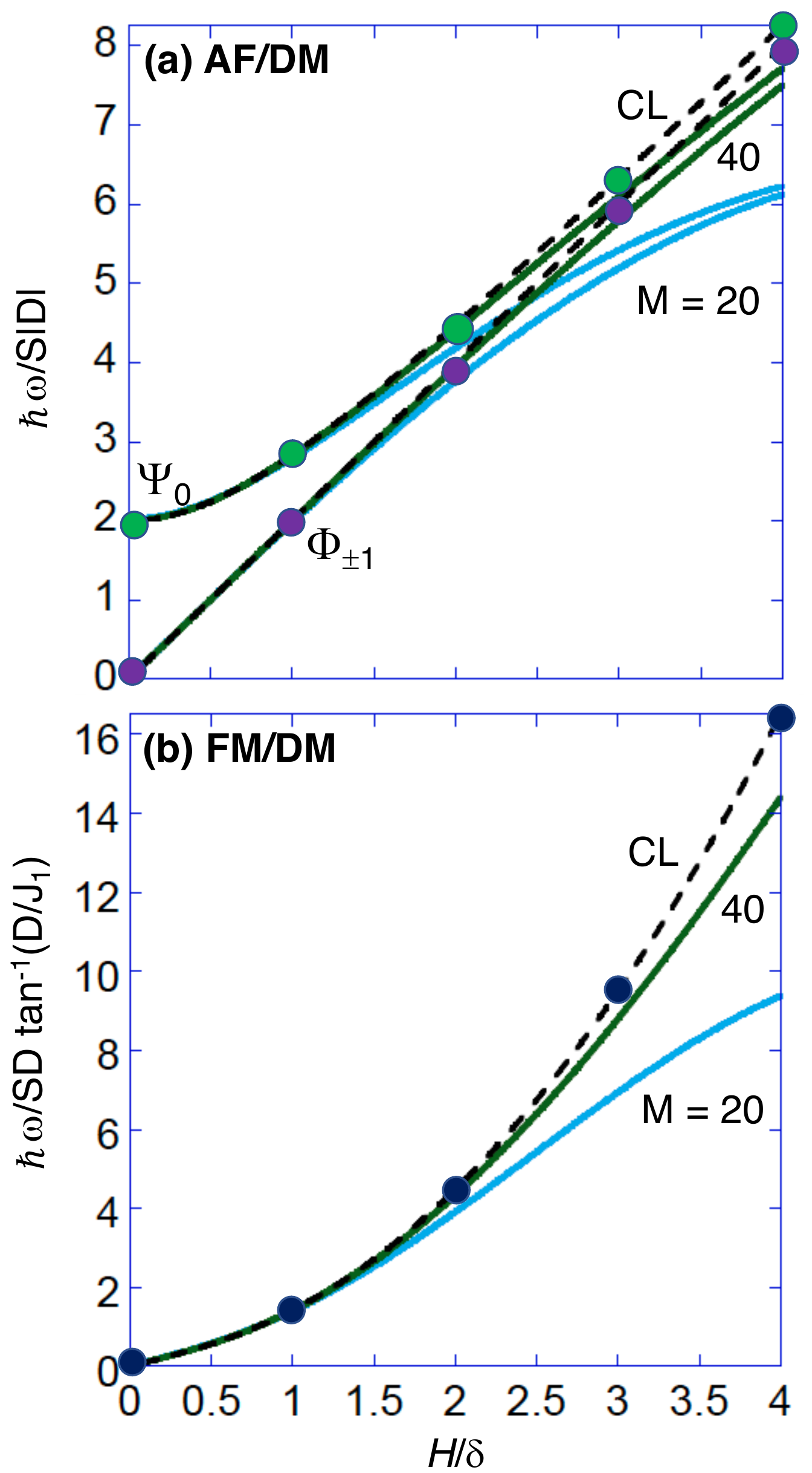}
\caption{(Color online)  A comparison of the scaled SW frequencies (a) $\hbar \omega /S\vert D\vert $ (AF/DM)
or (b) $\hbar \omega /SD \tan^{-1}(D/J_1)$ (FM/DM)
versus $H/\delta $ with $M=20$ (blue) and $40$ (green), 
and in the CL (dashed).  For the CL, normal modes at integer $n=H/\delta $
are denoted by closed circles.}
\end{figure}

Performing the same procedure for the exchange interaction carried out to second order in $\bm{v}$, 
we find ${\cal H}_{{\rm EX}}=\int d^dr \, h_{{\rm EX}}$ with
\begin{eqnarray}
h_{{\rm EX}}&=&
-\frac{v_0 J_1}{2(g\mu_B)^2} \sum_{\bm{v}} \biggl\{ \vert \vM \vert^2 + \vM \cdot(\bm{v}\cdot\bm{\nabla})\vM \nonumber\\
&+&\frac{1}{2}\sum_{\alpha}\left[\bm{v}\cdot\bm{\nabla}M_{\alpha}\,\bm{v}\cdot\bm{\nabla}M_{\alpha}-\vert \bm{v}\cdot\bm{\nabla}M_{\alpha}\vert^2
\right] \biggr\} \nonumber\\
&=& -\frac{A'}{2}\vert \vM \vert^2\nonumber \\
&+&\frac{J'}{2} \biggl\{ \sum_{\alpha }\vert \bm{\nabla}M_{\alpha}\vert^2 - \nabla^2 \vert \vM \vert^2 \biggr\} .
\end{eqnarray}
The CL parameters are $A'=a^d J_1/(g\mu_B)^2$ and $J'=a^{d+2}J_1/(g\mu_B)^2$. 
For an infinite system, the $\nabla^2 \vert \vM \vert^2$ term can be integrated out.

In the Appendix, these CL expressions are used to derive the ground-state helical ($D'_1 \ne 0$, $D'_2 = 0$) and 
cycloidal ($D'_1 = 0$, $D'_2 \ne 0$) states.  Those states 
are identical to those found in Section II.   The excitations of the cycloidal or helical state 
propagate according to the Landau-Lifshitz equation of motion,
\begin{equation}
\frac{\partial\vM }{\partial t}=\gamma \vM\times \frac{\delta h}{\delta\vM}
\label{landau_lifshitzEOM}
\end{equation}
with gyromagnetic ratio $\gamma=-g\mu_B/\hbar < 0$.   
We consider small deviations $\Delta \vM =\vM - \vM_0$ from the ground state
by dropping contributions that are quadratic or higher order in $\Delta\vM $. 
Because $\delta h/\delta \vM \vert_{\vM=\vM_0}= 0$,
$\delta h/\delta (\Delta\vM )$ is linear in $\Delta\vM $ and
\begin{equation}
\frac{\partial\Delta\vM }{\partial t}=\gamma \vM_0\times \frac{\delta h}{\delta (\Delta\vM )}. 
\label{landau_lifshitzLinearizedEOM}
\end{equation}
Consequently, $\vM_0\cdot\Delta\vM $ is a constant of the motion and the
time dependence of $\Delta\vM$ must be perpendicular to $\vM_0$. 

As expected from our numerical results, 
the SW frequencies are identical for the helical and cycloidal states.  More remarkably, the mode frequencies are independent
of dimension $d$ and given by
\begin{equation}
\hbar \omega (\Theta_{\pm n}) = \frac{S{D_1}^2}{J_1} \vert n \vert \sqrt{1 + n^2}.
\end{equation}
The CL result $\hbar \omega J_1 /S{D_1}^2 = \vert n\vert \sqrt{1+n^2}$ is compared with numerical results for
$\hbar \omega /SD \tan^{-1} (D/J_1)$
in Fig.4(b).  The difference between the analytic and numerical results disappears
as $M\rightarrow \infty $.  

Generally, the CL results for the SW dispersions $\omega_n (\vq )$ depend on the orientation of $\vq $ and agree
with numerical results only when $\vq $ is along $\vx $.  In other directions, the SWs are softer.
This effect was also found in the AF/DM case \cite{desousa08}.

The CL results for the SW amplitudes in the AF/DM and FM/DM cases are described in the next section.
We shall see that the numerical results for a cycloid or helix of finite length approach those results as $\delta \rightarrow 0$
or $M\rightarrow \infty $.
For the AF/CE and FM/CE cases, 
CL calculations would require keeping the second-order derivatives of the magnetization density and have not been 
performed.

\section{SW amplitudes for finite length}

The spin oscillation
$\Delta \vS^{(n)}_r(\vq ,t)$ at site $r$ produced by SW mode $n$ with wavevector $\vq $
is generally given by \cite{Cohen, fishmanbook}
\begin{equation}
\Delta \vS^{(n)}_r(\vq, t) = 2\sqrt{N} \, {\rm Re} \Bigl\{ e^{-i\omega_n t} \delta \vS_r(n, \vq ) \Bigr\},
\end{equation}
\begin{equation}
\delta \vS_r(n, \vq ) = \langle 0 \vert \vS_r \vert n, \vq \rangle,
\end{equation}
where $\vert 0\rangle $ is the ground state, $\vert n,\vq \rangle $ is an excited state containing a single SW with energy $\omega_n(\vq )$ at
wavevector $\vq $, and $\vS_r$ is the quantum spin operator at site $r$.  Like the SW frequency $\omega_n(\vq )$, 
the SW amplitude $\delta \vS_r(n, \vq )$ is the same at wavevectors $\vq=m\vQ $ for any integer multiple $m$ (including 0) of $\vQ$.  

A close examination of the SW amplitudes for DM and CE cycloids or helices with AF or FM interactions reveals that 
\begin{eqnarray}
\delta \vS_r(\Phi_{\pm n}) &=& \Bigl\{ \xi_1^{(n)} {\bf t}_r (-1)^r -i \xi_2^{(n)} \vp \Bigr\} e^{\pm 2\pi i n \delta r}\hskip .06cm {\rm (AF)},\hskip .65cm
\label{phin} \\
\delta \vS_r(\Psi_{\pm n})&= &\Bigl\{ \rho_1^{(n)} \vp (-1)^r +i \rho_2^{(n)} {\bf t}_r \Bigr\} e^{\pm 2\pi i n \delta r}\hskip .06cm {\rm (AF)},\hskip .65cm
\label{psin} \\
\delta \vS_r(\Theta_{\pm n}) &=&
 \Bigl\{ \gamma_1^{(n)} {\bf t}_r - i\gamma_2^{(n)} \vp  \Bigr\} e^{\pm 2\pi i n \delta r}\hskip 0.84cm {\rm (FM)},
\label{thin}
\end{eqnarray}
for either $q=0$ or $q=Q$.   As discussed above, the transverse direction $\vp $ is $\vy $ for a cycloid and $-\vx $ for a helix.
In each case, the real and positive coefficients are the same for the degenerate $\pm n$ modes and are normalized by taking
$\xi_1^{(n)2}+\xi_2^{(n)2}=1$, $\rho_1^{(n)2}+\rho_2^{(n)2}=1$, and $\gamma_1^{(n)2}+\gamma_2^{(n)2}=1$.  
The complex factors in the brackets imply that the tangential and transverse spin oscillations are out of phase.

As seen below, the SW amplitudes for the Goldstone modes are purely transverse (out of the classical-spin plane) or tangential 
(in the classical-spin plane).  For AF interactions, 
$\delta \vS_r(\Phi_0)= {\bf t}_r (-1)^r$ in both the DM and CE cases.   
In the AF/CE case, $\delta \vS_r(\Psi_{\pm 1}) = \exp (\pm 2\pi i\delta r)(-1)^r\,\vp $.
For FM interactions, $\delta \vS_r(\Theta_0)={\bf t}_r$ in both the DM and CE cases.
In the FM/CE case, $\delta \vS_r(\Theta_{\pm 1}) = \exp (\pm 2\pi i\delta r)\,\vp $.
Although not a Goldstone mode, the SW amplitude $\delta \vS_r(\Psi_0) = (-1)^r\,\vp $ of the AF/CE mode $\Psi_0$ 
is purely transverse but out of phase with the cycloid or helix.

\begin{figure}
\includegraphics[width=8.5cm]{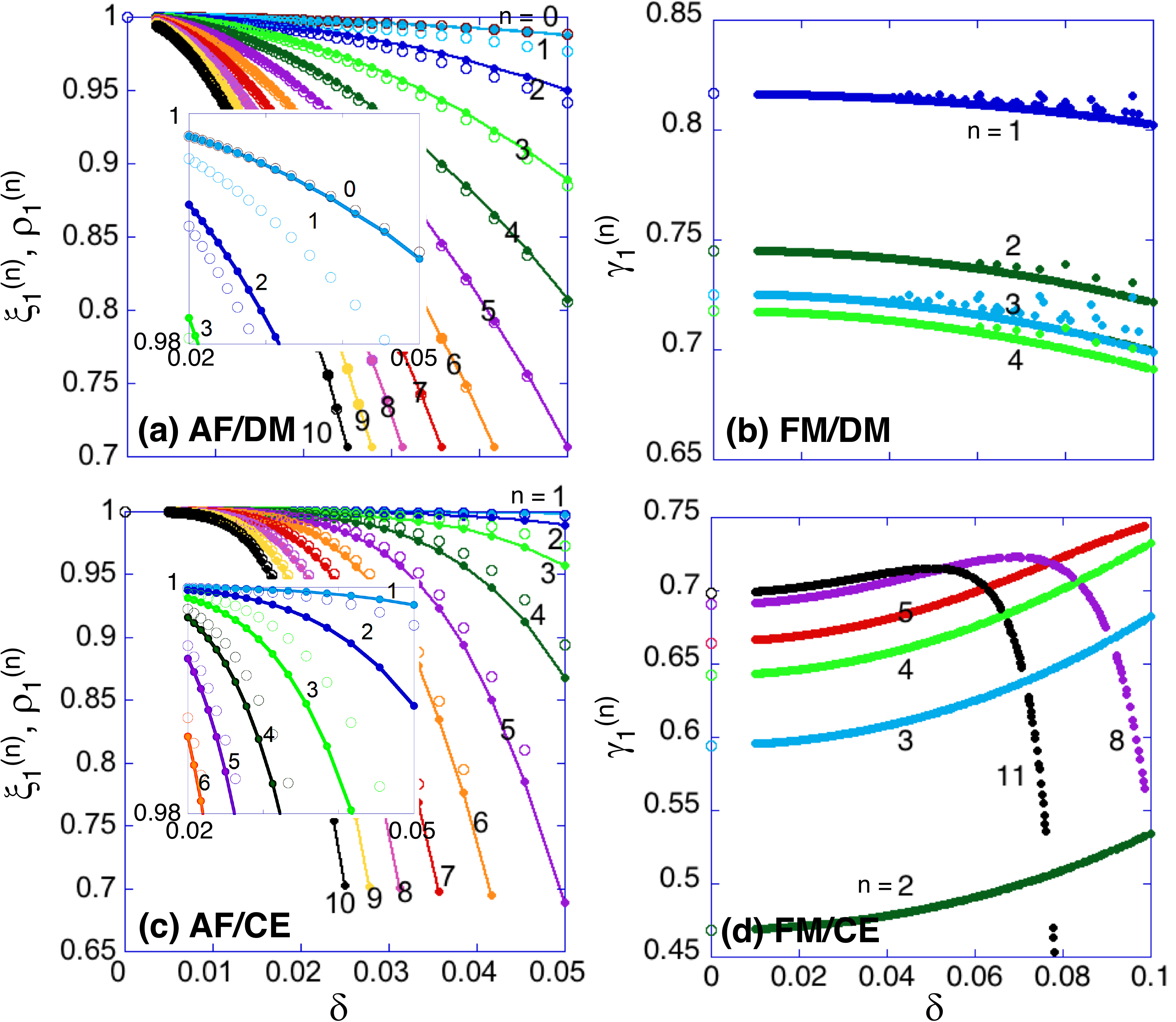}
\caption{(Color online)  Coefficients $\xi_1^{(n)}$ and $\rho_1^{(n)}$ (AF) or $\gamma_1^{(n)}$ (FM) 
versus $\delta $ for the same four cases as in Fig.2.  In (a) and (c), closed circles give $\xi_1^{(n)}$ and open circles give
$\rho_1^{(n)}$.  Points at $\delta =0$ are exact results in the CL for the FM/DM case and extrapolations
for the FM/CE case.  Goldstone modes not shown.}
\end{figure}

While even and odd $M$ were handled differently for AF interactions, physical results only depend on the wavevector parameter $\delta $.
The amplitude coefficients are plotted versus $\delta $ in Fig.5.  In either the AF/DM or AF/CE case, 
$\xi_1^{(n)}$ and $\rho_1^{(n)}$ approach
1 for all $n$ in the CL $\delta \rightarrow 0$.   For the AF/DM case, the CL
results $\xi_1^{(n)}=1$ and $\rho_1^{(n)}=1$ were obtained by de Sousa and Moore \cite{desousa08}.
Although the SW amplitudes become purely tangential or transverse as $\delta \rightarrow 0$,
the coefficients with larger $n$ converge much more slowly than for smaller $n$.
Figures 5(a) and (c) plot $\xi_1^{(n)}$ and $\rho_1^{(n)}$ as closed and open circles, respectively.  For larger $n$, $\xi_1^{(n)}$ and $\rho_1^{(n)}$
are quite close, but deviations can be seen for smaller $n$ away from $\delta =0$.

For FM interactions, the behavior of the coefficients is more complex.  While $\gamma_1^{(n)}\rightarrow 1/\sqrt{2}$ 
as $\delta \rightarrow 0$ and $n \rightarrow \infty $
in both the FM/DM and FM/CE cases, $\gamma_1^{(n)}$ have higher (FM/DM) or lower (FM/CE) limits for smaller $n>0$. 
Recall that $\gamma_1^{(1)}=0$ for the FM/CE case while $\gamma_1^{(0)}=1$ for both FM cases.
In the CL of the FM/DM case, the method discussed in the previous section reveals that 
\begin{eqnarray}
&\gamma_1^{(n)} &\rightarrow \sqrt{\frac{1+n^2}{1+2n^2}},\\
&\gamma_2^{(n)}&\rightarrow \frac{\vert n\vert }{\sqrt{1+2n^2}}.
\label{CLFM}
\end{eqnarray}  
Although not a rigorous proof,
we numerically find that 
\begin{eqnarray}
&\gamma_1^{(n)}&\rightarrow \frac{ \vert n^2-1\vert }{ \sqrt{2n^4+2n^2+1}},\\ 
&\gamma_2^{(n)}&\rightarrow \vert n\vert \sqrt{\frac{n^2+4}{2n^4+2n^2+1}}
\end{eqnarray}
in the CL of the FM/CE case.
So non-Goldstone modes always mix tangential and transverse components for FM interactions.

\begin{figure}
\includegraphics[width=8.5cm]{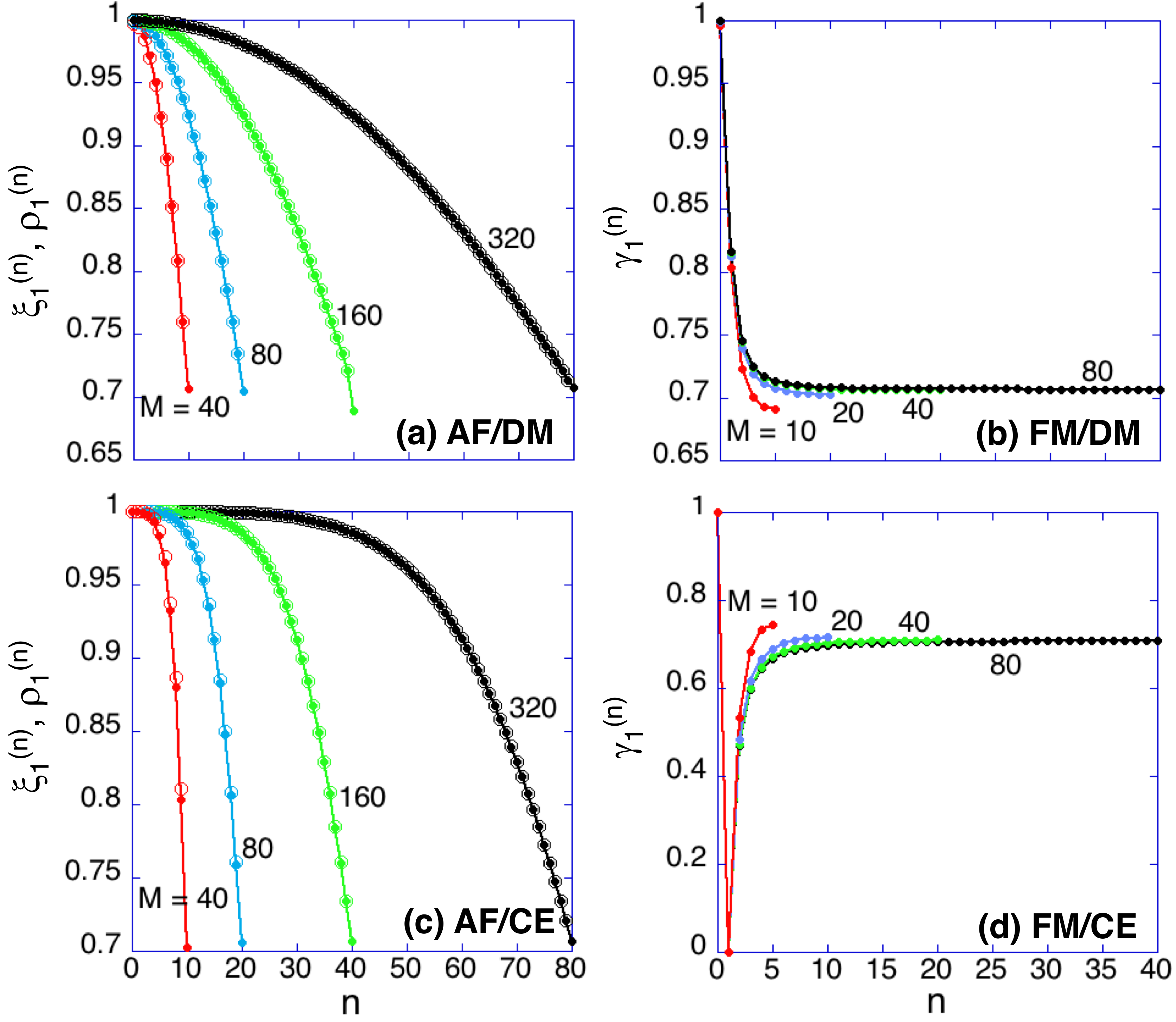}
\caption{(Color online)  Coefficients $\xi_1^{(n)}$ and $\rho_1^{(n)}$ (AF) or $\gamma_1^{(n)}$ (FM) versus
mode index $n $ for $\delta =1/M$ ($p=2$ for AF interactions and $p=1$ for
FM interactions) 
and the same four cases as in Fig.2.  In (a) and (c), closed circles give $\xi_1^{(n)}$ and open circles give
$\rho_1^{(n)}$.  The maximum $n$ is $M/4$ for AF interactions and $M/2$ for FM interactions.
}
\end{figure}
 
Another way to look at these results is by plotting the coefficients versus $n$ for a fixed $\delta = 1/M$ in Fig.6.
For AF interactions, the coefficients quickly fall off from their asymptotic $\delta \rightarrow 0$ limits of $\xi_1^{(n)}= 1$
and $\rho_1^{(n)} =1$ with increasing $n$.
As in Fig.5, the results for $\xi_1^{(n)}$ (closed circles) and $\rho_1^{(n)}$ (open circles) are very close.
Figures 5 and 6 suggest that for the maximum $n=M/4$, $\xi_1^{(n)}$ and $\rho_1^{(n)}$ approach $1/\sqrt{2}$ as $M$ increases.
In the FM/DM case, $\gamma_1^{(n)}$ falls off monotonically with $n$ for all $M$ and analytic results in the CL are indistinguishable
from numerical results for $M=80$.
In the FM/CE case, $\gamma_1^{(n)}$ increases with $n$ starting with $\gamma_1^{(1)}=0$.
In both FM cases, $\gamma_1^{(n)}$ remains fairly constant as a function of $n$ beyond $n=10$ or so and 
approaches $1/\sqrt{2}$ for large $M$.

\section{Observing the normal modes}

What do these results imply about the observability of the SW modes?  The contribution of mode $n$ to the
spectral weight $S_{\alpha \beta }(Q,\omega_n)$ is proportional to \cite{fishmanbook}
\begin{equation}
\sum_{r,s=1}^M  e^{-iQ(r-s)a } \delta S_{r \alpha }(n) \, \delta S_{s\beta }(n)^{\star }.
\end{equation}
Using Eqs.~(\ref{phin}-\ref{thin}) with $Q>0$, 
it is straightforward to show that the three modes $\Phi_0$ ($\alpha $ and $\beta $ tangential), $\Psi_1$ ($\alpha $ and $\beta $ transverse), and $\Phi_2$ ($\alpha $ and $\beta $ tangential) contribute
for AF interactions while the three modes 
$\Theta_0$ ($\alpha $ and $\beta $ tangential), $\Theta_1$ ($\alpha $ and $\beta $ transverse), and $\Theta_2$ ($\alpha $ and $\beta $ tangential) contribute for FM interactions.  
These modes are responsible for the INS intensity \cite{ins}
$S(q,\omega )= S_{yy}(q,\omega )+S_{zz}(q,\omega )$
plotted in Fig.1.

\begin{table*}
\caption{Observable modes with INS or THz spectroscopy}
%\begin{ruledtabular}
\begin{tabular}{|cc|cc|cc|cc|cc|}
\hline
 && AF/DM && FM/DM && AF/CE && FM/CE &\\
\hline
$S(Q,\omega )$ && $\Phi_0$, $\Psi_1$, $\Phi_2$  && $\Theta_0$, $\Theta_1$, $\Theta_2$ && $\Phi_0$, $\Psi_1$, $\Phi_2$  && $\Theta_0$, $\Theta_1$, $\Theta_2$  & \\
$\alpha^{\rm mag}(\omega )$ ($\delta > 0$) &&  $\Psi_{\pm 1}$  && $\Theta_{\pm 1}$   &&  ---  && --- & \\
$\alpha^{\rm mag}(\omega )$ (CL) &&  ---  && $\Theta_{\pm 1}$   && ---  &&  --- &\\
Examples:   && BiFeO$_3$ [\onlinecite{sosnowska95}], && MnSi  [\onlinecite{Bak80, kugler15, Shanavas16}]  &&  MnWO$_4$  [\onlinecite{arkenbout06}], && Sr$_3$Fe$_2$O$_7$ [\onlinecite{kim14}] & \\
 &&  Ba$_2$CuGe$_2$O$_7$ [\onlinecite{zheludev96}]  &&    && Ni$_3$V$_2$O$_8$ [\onlinecite{lawes05}]   &&  &\\
  \hline
\end{tabular}
%\end{ruledtabular}
\label{obmodes}
\end{table*}

The purely magnetic contribution of mode $n$ to the optical absorption $\alpha (\omega )$ is proportional to \cite{fishmanbook}
\begin{equation}
\frac{\omega_n}{(g\mB )^2}\, \Bigl\vert \langle 0 \vert {\bf h}\cdot \vM \vert n, q=0 \rangle \Bigr\vert^2 = \omega_n\, \Biggl\vert \sum_{r=1}^M {\bf h}\cdot \delta \vS_r(n) \Biggr\vert^2 ,
\end{equation}
where ${\bf h}$ is the magnetic polarization of light and $\vM = -g\mB \sum_{r=1}^M \vS_r$ is the magnetization per unit cell.  This is
nonzero for $\Psi_{\pm 1}$ in the AF/DM case and for $\Theta_{\pm 1}$ in the FM/DM case, both when ${\bf h}$ is in the classical-spin plane
of the 
cycloid or helix.
So for nonzero $\delta $, optical spectroscopy will detect two modes ($\Psi_{\pm 1}$) in the AF/DM case,
two ($\Theta_{\pm 1}$) in the FM/DM case, and none in the CE cases.
Only the FM/DM $\Theta_{\pm 1}$ modes remain optically active as $\delta \rightarrow 0$.  

Modes that are observable by INS or THz spectroscopy
are summarized in Table \ref{obmodes}.  Notice that different parts of $\delta \vS_r(n)$ contribute to the INS intensity and 
to the optical absorption.
For the AF/DM $\Psi_{\pm 1}$ and FM/DM $\Theta_{\pm 1}$ modes, the tangential parts of $\delta \vS_r(n)$ contribute to the optical absorption 
while the transverse parts contribute to the INS intensity.

It is tempting to argue that modes with no spectral $S(Q,\omega )$ or magnetic optical $\alpha^{\rm mag} (\omega )$ weight
are not physically significant but 
rather artifacts of our numerical solutions for ${\cal H}_{\rm DM}$ and ${\cal H}_{\rm CE}$.  However, those modes 
are eigenstates of the Hamiltonian with nonzero eigenvectors and well-defined SW amplitudes.  Rather than
a trivial consequence of zone folding, all $M$ modes are required by the $M$ degrees of freedom in the magnetic unit cell
of the cycloid or helix.  

Up to some maximum value \cite{maxn} for the mode number $n$, 
all predicted modes appear in the spectral weight $S(q,\omega )$ at some multiple of $H=qa/2\pi = \delta $.  
Consider, for example, the spectra in Fig.2 with $\delta = 0.1$.  For AF interactions, $\Phi_1$ has no
spectral weight at $H=0$ or $0.4$ but gains spectral weight at $H=0.1$ in Figs.2(a) and (c).   For FM interactions, $\Theta_3$
appears in the spectral weight of Figs.2(b) and (c) at $H= 0.3$.  

Although only a handful of modes contribute to the magnetic optical absorption $\alpha^{\rm mag}(\omega )$, 
the optical weight of the ``hidden" modes can be switched on by several physical perturbations \cite{caveat}
that do not significantly alter their frequencies.  
For the AF/DM compound BiFeO$_3$, 
easy-axis anisotropy \cite{sosnowska95} makes $\xi_2^{(0)}$ nonzero
so that $\Phi_0$ (no longer a Goldstone mode) becomes optically active for ${\bf h}=\vy $.  Hybridization with $\Phi_0$ then activates \cite{fishman12} $\Phi_{\pm 2}$,
also for ${\bf h}=\vy $.
The alternating tilt of the cycloid \cite{kadomtseva04} 
on neighboring hexagonal planes mixes transverse and tangential components,
thereby activating \cite{fishman13} $\Psi_0$ and $\Phi_{\pm 1}$.  Consequently, eight modes (four accounting for their degeneracies
and excluding the low-frequency mode $\Phi_0$)
appear in the THz \cite{talbayev11} spectra of BiFeO$_3$ in zero field.   
Due to hybridization, a magnetic field activates the complete mode spectrum \cite{nagel13} with frequencies that
nicely extrapolate to the frequencies of the zero-field ``hidden" modes.
While the selection rules for the Raman spectra
are more complex than for the THz spectra, all of the predicted spectroscopic modes $\Phi_n$ and $\Psi_n$ 
seem to appear in the Raman \cite{caz08} spectra of BiFeO$_3$.

$\,\, $

\section{Conclusion}

How do other well-known materials with cycloidal or helical states fall into the four cases considered here?
Along with BiFeO$_3$, the multiferroic Ba$_2$CuGe$_2$O$_7$ is also a member of the AF/DM class \cite{zheludev96}.
With helical AF2 and cycloidal AF5
states created by long-range competing AF interactions \cite{arkenbout06}, Co-doped \cite{ye12} MnWO$_4$ falls into the AF/CE class.
So do the cross-tie spins of Ni$_3$V$_2$O$_8$ in its low-temperature C$^{\prime }$ phase \cite{lawes05, Ehlers13}.
Although itinerant \cite{ishikawa76}, MnSi is a member of the FM/DM family \cite{Bak80, Shanavas16} and its
inelastic neutron-scattering spectra \cite{kugler15} agrees well with Fig.2(b).  Of the three observed modes in MnSi, only the 
central $\Theta_1$ mode is predicted to be optically active.
A rare member of the FM/CE class, Sr$_3$Fe$_2$O$_7$ has a helical state produced by the competition between 
FM nearest-neighbor double exchange and AF next-nearest neighbor exchange \cite{kim14}.
However, A-type AF materials with FM nearest-neighbor interactions and CE within a plane 
might also be described by our FM/CE results. 

To summarize, we have evaluated the normal modes of a spin cycloid or helix produced by either DM or CE interactions
and for either AF or FM nearest-neighbor exchange coupling.   
In the CL for AF exchange, the SW amplitudes for all modes are either purely tangential or transverse.
But for FM exchange, the SW amplitudes for all modes except the Goldstone modes contain both 
tangential and transverse components, even in the CL.  Whereas the mode spectrum for DM interactions contains only one
Goldstone mode, the mode spectrum for CE interactions contains three Goldstone modes.
Our results explain why only a subset of these modes are observed using neutron scattering or optical absorption.

Research by RF sponsored by the U.S. Department of Energy, 
Office of Basic Energy Sciences, Materials Sciences and Engineering Division.
TR would like to acknowledge support from the Estonian Ministry of Education and Research with institutional research funding IUT23-3, and the European Regional 
Development Fund Project No. TK134.
RdS acknowledges financial support from NSERC (Canada) through its
Discovery program (RGPIN-2015-03938).

\appendix

\section{CL and generalization to higher dimensions for the FM/DM case}

In this appendix, we compute the ground state and excitation spectra of a spiral magnet using continuum field theory.  
Results are valid when $Qa,qa\ll 1$, where $Q$ and $q$ are the ground state spiral wavevector and excitation (SW) wavevector, respectively, and $a$ is the lattice parameter. In this regime, the discrete lattice calculations approach the CL results. 

\subsection{Ground state in the continuum limit}

The CL of the Hamiltonian was derived in Section IV above.
Combining the results from that section, we get the total Hamiltonian density
\begin{eqnarray}
h&=&-\frac{A'}{2}\left|\bM \right|^2+\frac{J'}{2} \sum_{\alpha }\left|\bm{\nabla}M_{\alpha}\right|^2 +\frac{V'}{4} \left|\bM \right|^4
\nonumber\\
&-&D'_{1}\, \bM \cdot \left(\bm{\nabla}\times\bM \right)+D'_{2}\, \zz \cdot\Bigl\{ \bM \left(\bm{\nabla}\cdot\bM \right)
\nonumber \\ 
&+&\bM \times \left(\bm{\nabla}\times \bM \right)\Bigr\} ,
\label{hdensfinal}
\end{eqnarray}
with the direction $\ee $ of the ferroelectric moment taken along $\zz $. 
The positive contribution $V' M^4/4$ imposes a smooth bound on $M$ and allows 
us to determine a ground-state function $\bM _0(\br )$ that satisfies the stationarity condition, 
\begin{equation}
\frac{\delta h}{\delta \bM }\bigg|_{\bM =\bM_0}= 0.
\label{dhdm0}
\end{equation}
Of course, all final results must be independent of $V'$.   

The functional derivative is given by 
\begin{eqnarray}
\frac{\delta h}{\delta \bM }&=& \frac{\partial h}{\partial \bM }-\bm{\nabla}\cdot\frac{\partial h}{\partial \bm{\nabla}\bM }\nonumber\\
&=&\left(-A'+ V'M^2\right)\bM -J'\nabla^2\bM\nonumber\\
&-&2D'_1 \left(\bm{\nabla}\times \bM\right)-2D'_2 \left(\zz \times \bm{\nabla}\right)\times\bM.
\label{dhdm}
\end{eqnarray}
We focus on the family of \emph{harmonic spiral states}:
\begin{eqnarray}
\bM_0(\br )&=&\frac{1}{2}\Bigl\{ \blm^* \, e^{i\vQ \cdot\br } +\blm \, e^{-i\vQ \cdot\br } \Bigr\} \nonumber\\
&=&\cos (\vQ \cdot\br ) \bcr+\sin (\vQ\cdot\br) \bci ,
\label{harmspir}
\end{eqnarray}
where $\blm=\bcr+i\bci$ and $\blm_{{\rm R,I}}$ are real vectors. These states satisfy $\nabla^2 \bM_0=-Q^2\,\bM_0$. 
To connect with the numerical results for $D >0$, we take $D_1' >0$ and $D_2' < 0$.

\subsubsection{Helix:  $D'_1> 0, D'_2=0$} 

Based on Eq.~(\ref{dhdm}), the functional derivative vanishes provided that $\bm{\nabla}\times \bM_0$ equals a 
constant times $\bM_0$. 
Since 
\begin{equation}
\bm{\nabla}\times \bM_0 = -\vQ \times \Bigl\{ \bcr \sin (\vQ \cdot\br) 
- \bci \cos (\vQ\cdot\br) \Bigr\},
\end{equation}
choosing $\vQ $, $\bci $, and $\bcr $ to be a set of mutually orthogonal vectors does the job. 
Such a state is called a \emph{circular helix} because $\vQ $ is perpendicular to $\bM_0$ at all points in space. 
Since the orientation of $\vQ $ is arbitrary, we can pick $\vQ =Q\xx $ without loss of generality. In this case,
\begin{equation}
\bM_0(\br)=M_0 \bigl( 0, \sin{(Qx)}, \cos{(Qx)}\bigr).
\label{m0helix}
\end{equation}
Note that $\bm{\nabla}\times \bM_0=Q\,\bM_0$ and $\bm{\nabla}\cdot \bM_0=0$. 
Using this function in Eq.~(\ref{dhdm}), we obtain
\begin{equation}
{M_0}^{2}=\frac{A'-J'Q^2+2D'_1 Q}{V'},
\label{m0sq}
\end{equation}
where $V'$ can be adjusted to obtain the desired maximum spin. 
Minimizing
\begin{equation}
h=-\frac{A'}{2}{M_0}^{2}+\frac{J'}{2}Q^2{M_0}^{2}+\frac{V'}{4}{M_0}^{4}-D'_{1}Q{M_0}^{2}, 
\end{equation}
with respect to $Q$ gives
\begin{equation}
Q_{{\rm helix}}=\frac{D'_1}{J'}= \frac{D_1}{J_1},
\label{qhelix}
\end{equation}
which is the CL limit of Eq.~(\ref{DJ1}) with $D = D_1 > 0$.

\subsubsection{Cycloid:  $D'_1 = 0, D'_2< 0$} 

Following the same procedure, a family of local energy minima can be found by imposing the condition
$\left(\zz \times \bm{\nabla}\right)\times\bM_0 \propto \bM_0$ in Eq.~(\ref{dhdm}). From Eq.~(\ref{harmspir}) we get
\begin{eqnarray}
\left(\zz \times \bm{\nabla}\right)\times\bM_0 &=& -\left(\zz \times\vQ \right)\times \Bigl\{
\bcr \sin (\vQ \cdot\br) \nonumber\\
&-&\bci \cos (\vQ \cdot\br)\Bigr\}.
\end{eqnarray}
Choose $\zz \times\vQ$, $\bci $, and $\bcr $ to be a set of mutually orthogonal vectors with $\vQ \perp \zz $.  Either 
$\bcr \parallel\zz $ and $\bci \parallel\vQ $ or $\bcr \parallel\vQ $ and $\bci \parallel\zz $. 
This leads to the \emph{cycloidal} state with $\vQ = Q\xx $:
\begin{equation}
\bM_{0}(\br)=M_0 \bigl(\sin (Qx),0 ,\cos(Qx )\bigr),
\label{m0cycloid}
\end{equation}
which satisfies $\left(\zz \times \bm{\nabla}\right)\times\bM_0 = -Q\bM_0$. 
Eq.~(\ref{dhdm}) then implies that
\begin{equation}
{M_0}^{2}=\frac{A'-J'Q^2+2\vert D'_2\vert  Q}{V'}. 
\label{m0sqcycloid}
\end{equation}
Minimizing
\begin{equation}
h=-\frac{A'}{2}{M_0}^{2}+\frac{J'}{2}Q^2{M_0}^{2}+\frac{V'}{4}{M_0}^{4}-\vert D'_{2}\vert Q{M_0}^{2}.
\end{equation}
with respect to $Q$ gives 
\begin{equation}
Q_{{\rm cycloid}}=\frac{\vert D'_2\vert }{J'}= \frac{\vert D_2\vert }{J_1},
\label{qcycloid}
\end{equation}
which is the CL limit of Eq.~(\ref{DJ1}) with $D = -D_2 > 0$.

\subsection{SW excitations in the CL}

The excitations of the cycloidal or helical
state propagate according to the Landau-Lifshitz equation of motion given by Eq.~(\ref{landau_lifshitzEOM}).  We now separately
consider the helical and cycloidal SWs.

\subsubsection{Helical SWs}

The linear excitations of the helical state can be parametrized as 
\begin{equation}
\Delta\bM =\psi \, \xx +\phi \, \tt (x), 
\end{equation}
where $\tt (x)=\bigl(0,-\cos (Qx),\sin (Qx)\bigr)$ is the unit vector tangential to the helix.  Compared to the tangent
in Eq.~(\ref{tnghl}), the change in sign in $\tt (x)$ is required because $M_0 < 0$ in Eq.~(\ref{m0helix}) for $\vM_0 ({\bf r})$. 
Plugging this into Eq.~(\ref{dhdm}) with $D'_2=0$ and using Eqs.~(\ref{m0sq})~and~(\ref{qhelix}) gives
\begin{eqnarray}
\frac{\delta h}{\delta (\Delta\bM )}&=& J'\xx \Bigl\{ Q^2 -\nabla^2\Bigr\}\psi  +J' \tt (x) \, \nabla^2 \phi  \nonumber\\
&+&2D'_1 \Bigl\{  \xx \times \bm{\nabla}\psi -\tt (x) \times \bm{\nabla}\phi \Bigr\} ,
\end{eqnarray}
with $Q=D'_1/J'$.  

Resolving Eq.~(\ref{landau_lifshitzLinearizedEOM}) into components along $\xx$ 
and $\tt (x)$ produces the coupled differential equations
\begin{eqnarray}
\partial_t \psi &=& \gamma M_0 J'\Bigl\{
\nabla^2 \phi  +2 Q \Bigl[ \sin (Qx) \, \partial_y \psi \nonumber \\
&+&\cos (Qx) \, \partial_z \psi \Bigr] \Bigr\} ,\\
\partial_t \phi &=& -\gamma M_0 J'\Bigl\{
\Bigl[ \nabla^2- Q^2\Bigr] \psi - 2Q 
\nonumber\\
&\times &\Bigl[\sin (Qx) \, \partial_y \phi+\cos (Qx) \, \partial_z \phi \Bigr]\Bigr\}.
\label{helixqsystem}
\end{eqnarray}
Modes propagating with wavevector $\qq =q\xx $ are plane waves 
$\psi=\psi_0\exp \{i(qx-\omega t)\}$ and $\phi=\phi_0\exp \{i(qx-\omega t)\}$ satisfying the eigenvalue equation
\begin{equation}
\begin{pmatrix}
-i\omega /\gamma M_0 && J' q^2 \\
-J'(Q^2+q^2) && -i\omega/ \gamma M_0
\end{pmatrix}
\begin{pmatrix}
\psi_0 \\ \phi_0
\end{pmatrix}
= 0
\label{homhelix}
\end{equation}
with solutions
\begin{equation}
\omega(q)=\pm \gamma M_0D'_1\; q\, \sqrt{1+\left(\frac{q}{Q}\right)^2}.
\label{dispersion_helix}
\end{equation}
For small $q$, the helical magnons propagate linearly with $q$ like light, 
in contrast to the $q^2$ dispersion found in conventional FMs. 

With $r=x/a$, the fluctuation $\Delta\bM_r$ on site $r$ is given by 
\begin{equation}
\Delta \bM_r (q,t) = {\rm Re}\Bigl\{ \bigl( \psi_0 \, \xx+\phi_0 \, \tt (r) \bigr) e^{ i(qra-\omega t)} \Bigr\} 
\end{equation} 
where $(\psi_0,\phi_0)$ is the eigenvector of Eq.~(\ref{homhelix}).   Including both
$\omega (nQ) = \pm \omega_n$ solutions, we find $\Delta \bM_r (nQ,t) = \Delta \bM_r^{(\pm n)}(t)$ with
\begin{eqnarray}
&&\Delta\bM_r^{(\pm n)}(t)=\phi_0\biggl\{
\cos (\pm nQra-\omega_n t) \, \tt (r) \nonumber \\
&&-\frac{\vert n\vert }{\sqrt{1+n^2}}\sin (\pm nQ ra-\omega_n t)\, \xx
\biggr\}.
\end{eqnarray}

The solutions of Eq.~(\ref{helixqsystem}) with $\vq $ along the $y$ and $z$ directions are  
not simple plane waves. Rather, they are  
Bloch waves that mix integer multiples of $Q$, with dispersion $\omega\propto q^2$ when $\vq$ is perpendicular to $\xx $.
A similar effect was found for itinerant cubic magnets \cite{Maleyev06} and for cycloidal AFs \cite{desousa08}. 

\subsubsection{Cycloidal SWs}

The linear excitations of the cycloidal state can be written as
\begin{equation}
\Delta\bM=\psi \, \yy+\phi \, \tt (x),
\end{equation}
with $\tt (x)=\bigl(-\cos (Qx),0, \sin (Qx) \bigr)$ as the tangential unit vector. 
Compared to the tangent
in Eq.~(\ref{tngcy}), the change in sign in $\tt (x)$ is required because $M_0 < 0$ in Eq.~(\ref{m0cycloid}) for $\vM_0 ({\bf r})$. 
Plugging this into Eq.~(\ref{dhdm}) with $D'_1=0$ and using Eqs.~(\ref{m0sqcycloid})~and~(\ref{qcycloid}) gives 
\begin{eqnarray}
\frac{\delta h}{\delta (\Delta\bM )}&=& -J'\yy \Bigl\{ \Bigl[ \nabla^2-Q^2 \Bigr] \psi -2Q \sin (Qx)\, \partial_y \phi \Bigr\}\nonumber \\
&+&J' \tt (x) \Bigl\{ \nabla^2 \phi  -2Q \sin{(Qx)}\,\partial_y \psi \Bigr\},
\end{eqnarray}
with $Q=\vert D'_2\vert /J'$.  

Resolving Eq.~(\ref{landau_lifshitzLinearizedEOM}) into components along $\yy $ and $\tt (x)$
produces the coupled differential equations:
\begin{eqnarray}
\partial_t \psi &=& -\gamma M_0 J' \Bigl\{
\nabla^2 \phi
-2 Q \sin (Qx)\, \partial_y \psi \Bigr\} ,\\
\partial_t \phi &=& \gamma M_0 J' \Bigl\{ \Bigl[ \nabla^2 -Q^2\Bigr] \psi \nonumber \\
&-&2Q \sin (Qx) \, \partial_y \phi \Bigr\}.
\label{cycloidqsystem}
\end{eqnarray}
Modes propagating with ${\bf{q}} = q\xx $ are simple plane waves  
$\psi=\psi_0\exp \{i(qx-\omega t)\}$
 and $\phi=\phi_0\exp \{i(qx-\omega t)\}$ satisfying the eigenvalue equation
 \begin{equation}
\begin{pmatrix}
-i\omega/ \gamma M_0 && -J' q^2 \\
J'(Q^2+q^2) && -i\omega / \gamma M_0
\end{pmatrix}
\begin{pmatrix}
\psi_0 \\ \phi_0
\end{pmatrix}
= 0,
\label{homcycloid}
\end{equation}
which leads to the same dispersion obtained for the helix, Eq.~(\ref{dispersion_helix}),
with $D_2'$ replacing $D_1'$.
The expression for $\Delta\bM_r^{(\pm n)}(t)$ is also quite similar:
\begin{eqnarray}
&&\Delta \bM_r^{(\pm n)}(t)=\phi_0\biggl\{
\cos (\pm nQra-\omega_n t) \, \tt (r) \nonumber \\
&&+\frac{\vert n\vert }{\sqrt{1+n^2}}\sin (\pm nQ ra-\omega_n t)\, \yy .
\biggr\}.
\end{eqnarray}

$\,$

As for the helix, the solutions of Eq.~(\ref{cycloidqsystem}) propagating with a component of $\qq $ out of the $yz$ plane 
are more complex Bloch states that mix integer multiples of $Q$. 

$\,$

\subsection{Connection to discrete model and role of dimensionality}

Using $\gamma=-g\mu_B/\hbar$, $M_0=-g\mu_B S/a^d$, and $D'_{i}=a^{d+1}D_i/(g\mu_B)^2$,
we get the same SW dispersion for either the helix or the cycloid,
\begin{equation}
\hbar \omega (q)=\pm SD_i\, qa\sqrt{1+\left(\frac{q}{Q}\right)^2}.
\end{equation}
With $\omega (nQ) =\pm \omega_n$ and $Qa = D/J_1$,
\begin{equation}
\hbar \omega_n= \frac{SD^2}{J_1} \vert n \vert \, \sqrt{ 1+n^2}.
\end{equation}
This CL result is compared to numerical results for the $\Theta_{\pm n}$ modes with
finite $M$ in Fig.4(b). In the CL, neither the dispersion nor the eigenvector depends on dimensionality $d$. 
However, the dispersion in the CL changes when $\vq $ is not along $\vx $.

Normalizing $\Delta \bM_r^{(\pm n)} (t)$ for the $\Theta_{\pm n}$ modes with $Q=2\pi \delta /a $, we obtain
\begin{eqnarray}
&&\Delta \bM_r^{(\pm n)}(t) =\sqrt{\frac{1+n^2}{1+2n^2}}\cos (\pm 2\pi \delta n r-\omega_n t)\, \tt (x)\nonumber \\
&&+ \frac{\vert n\vert }{\sqrt{1+2n^2 }} \sin (\pm 2 \pi \delta n r -\omega_n t)\, \pp,\\
\,\nonumber
\end{eqnarray}
where $\pp = -\xx $ for the helix and $\yy $ for the cycloid.  Using the parameters in Eq.~(\ref{thin}), we find
\begin{eqnarray}
&\gamma_1^{(n)} &=\sqrt{\frac{1+n^2}{1+2n^2}},\\
&\gamma_2^{(n)} &= \frac{\vert n\vert }{\sqrt{1+2n^2}}.
\end{eqnarray}  
in agreement with our numerical results.

\vfill

\bibliographystyle{unsrt}

\begin{thebibliography}{10}
\bibitem{maxn}
The maximum $n$ is int($M/4$) for AF interactions and int($M/2$) for AF
  interactions.
\end{thebibliography}

\end{document}